\documentclass[amstex]{article}
\usepackage{graphicx}
\usepackage{dcolumn}
\usepackage{amsmath}
\usepackage{amssymb}
\usepackage{amsfonts}
\usepackage{amscd}
\begin{document}
\newtheorem{thm}{Theorem}[section]
\newtheorem{lem}{Lemma}[section]
\newtheorem{prop}{Proposition}[section]
\newtheorem{cor}{Corollary}[section]
\newtheorem{assum}{Assumption}[section]
\newtheorem{rem}{Remark}[section]
\newtheorem{defn}{Definition}[section]
\newcommand{\lv}{\left \vert}
\newcommand{\rv}{\right \vert}
\newcommand{\lV}{\left \Vert}
\newcommand{\rV}{\right \Vert}
\newcommand{\la}{\left \langle}
\newcommand{\ra}{\right \rangle}
\newcommand{\lcm}{\left [}
\newcommand{\rcm}{\right ]}
\newcommand{\ket}[1]{\lv #1 \ra}
\newcommand{\bra}[1]{\la #1 \rv}
\newcommand{\lmk}{\left (}
\newcommand{\rmk}{\right )}
\newcommand{\al}{{\mathfrak A}}
\newcommand{\md}{M_d({\mathbb C})}
\newcommand{\ali}[1]{{\mathfrak A}_{[ #1 ,\infty)}}
\newcommand{\alm}[1]{{\mathfrak A}_{(-\infty, #1 ]}}
\newcommand{\cb}{{\cal B}}
\newcommand{\aax}{a(\alpha)}
\newcommand{\ana}{a_n(\alpha)}
\newcommand{\tn}[1]{\lV\lv #1 \rv\rV_\theta}
\newcommand{\tdn}[1]{\lV\lv #1 \rv\rV_{\theta'}}
\newcommand{\ta}{{\cal B}\otimes \al_{[1,\infty)}}
\newcommand{\nn}[1]{\lV #1 \rV}
\newcommand{\pxn}{\lV\lv \Xi \rv\rV}
\newcommand{\psn}{\lV\lv \Psi\rv\rV}
\newcommand{\phn}{\lV\lv \Phi \rv\rV}
\newcommand{\et}{{\eta_t}}
\newcommand{\pt}{{\varphi_t}}
\newcommand{\en}{{\eta}}
\newcommand{\pn}{{\varphi}}
\newcommand{\dn}{\Delta_{\pn,\eta}}
\newcommand{\dt}{\Delta_{\pt,\et}}
\newcommand{\mpd}{{\cal M}_{*,+}}
\newcommand{\cm}{{\cal M}}
\newcommand{\ch}{{\cal H}}
\title{A generalization of
the inequality of Audenaert et al.}
\author{Yoshiko Ogata\thanks
{Graduate School of Mathematics, University of Tokyo, Japan}}
\maketitle{}
\begin{abstract}
We extend the inequality of
Audenaert et al \cite{fc} to 
general von Neumann algebras.
\end{abstract}
\section{Introduction}
Let $A,B$ be positive matrices and  $0\le s\le 1$.
Then an inequality 
\begin{align}\label{fci}
2Tr A^sB^{1-s}\ge Tr(A+B-\lv A-B\rv)
\end{align}
holds.
This is a key inequality to prove the upper bound of Chernoff bound,
in quantum hypothesis testing theory.
This inequality was first proven in \cite{fc}, using 
an integral representation of the function $t^s$.
Recently, N.Ozawa gave a much simpler proof for the same 
inequality.
In this note, based on his proof, we extend the inequality to 
general von Neumann algebras.
More precisely, we prove the following:
Let $\{\cm, \ch, J,{\cal P}\}$ be a standard form associated with
a von Neumann algebra $\cm$, i.e.,
$\ch$ is a Hilbert space where $\cm$ acts on,
$J$ is the modular conjugation, and $\cal P$ is the natural positive
cone.(See \cite{Takesaki})
Let ${\cal M}_{*+}$
be the set of all positive normal linear functionals
over $\cal M$.
For each $\pn\in {\cal M}_{*+}$, $\xi_\pn$
is the unique element in the natural positive cone $\cal P$
which satisfies $\pn(x)=(x\xi_{\pn},\xi_{\pn})$
for all $x\in{\cal M}$.
We denote the relative modular
 operator associated with
$\varphi,\psi\in {\cal M}_{*+}$
by $\Delta_{\pn\psi}$.(See Appendix.)
The main result in this note is the following:
\begin{prop}\label{main}
Let $\varphi,\eta$ be positive
normal linear functionals on a von Neumann algebra $\cal M$.
Then, for any $0\le s\le 1$, 
\begin{align}\label{mi}
\eta(1)-(\eta-\pn)_+(1)
\le 
\lV\Delta_{\en,\pn}^{\frac s2}\xi_{\pn}\rV^2.
\end{align} The equality holds iff
$\eta=(\en-\pn)_++\psi$ and $\pn=(\en-\pn)_-+\psi$
for some $\psi\in{\cal M}_{*+}$ whose support is orthogonal to the 
support of $\lv \en-\pn\rv$. 
\end{prop}
As a corollary of this proposition, we obtain
a generalization of the inequality of \cite{fc}:
\begin{cor}\label{fcc}
Let $\varphi,\eta$ be positive
normal linear functionals on a von Neumann algebra $\cal M$.
Then, for any $0\le s\le 1$, 
\begin{align}\label{mainineq}
2\lV\Delta_{\eta\varphi}^{\frac s2}\xi_{\pn}\rV^2
\ge
\varphi(1)+\eta(1)-\lv\varphi-\eta\rv(1).
\end{align}The equality holds iff
$\eta=(\en-\pn)_++\psi$ and $\pn=(\en-\pn)_-+\psi$
for some $\psi\in{\cal M}_{*+}$ whose support is orthogonal to the 
support of $\lv \en-\pn\rv$. 
\end{cor}
If $s=\frac 12$, this is the
Powers-St\o rmer inequality.
Applications of this inequality 
for hypothesis testing problem can
be found in \cite{jops}.
\section{Proof of Proposition \ref{main}}
We first prove the following lemma which
we need in the proof
of Proposition \ref{main}:
\begin{lem}\label{ichi}
Let $\varphi_1,\varphi_2,\psi,\eta$
be faithful normal positive linear functionals over a von Neumann algebra 
$\cal M$.
Assume that $\varphi_1\le\varphi_2$
and $\eta\le \psi$.
Then for all 
$0<s<1$,
\begin{align*}
\lV\Delta_{\varphi_2\eta}^{\frac s2}\xi_\eta\rV^2
-\lV\Delta_{\varphi_1\eta}^{\frac s2}\xi_\eta\rV^2
\le
\lV\Delta_{\varphi_2\psi}^{\frac s2}\xi_\psi\rV^2
-\lV\Delta_{\varphi_1\psi}^{\frac s2}\xi_\psi\rV^2.
\end{align*}
\end{lem}
{\it Proof}
First we consider the case $\varphi_2\le\psi$.
In this case, by Lemma \ref{cos}, 
$(D\varphi_1:D\psi)_t,(D\varphi_2:D\psi)_t,
(D\eta:D\psi)_t$ have continuations
$(D\varphi_1:D\psi)_z,(D\varphi_2:D\psi)_z,
(D\eta:D\psi)_z\in {\cal M}$, 
analytic on 
$I_{-\frac12}:=\{z\in{\mathbb C\;:\; -\frac12<\Im z<0}\}$
and bounded continuous on $\overline{I_{-\frac12}}$,
with norm less than or equal to $1$.\\
We define a positive operator
\begin{align}
T:=(D\varphi_2:D\psi)_{-i\frac s2}^*(D\varphi_2:D\psi)_{-i\frac s2}
-(D\varphi_1:D\psi)_{-i\frac s2}^*(D\varphi_1:D\psi)_{-i\frac s2}
\in {\cal M}.
\end{align}
To see that $T$ is positive, recall from Lemma \ref{cos}
that for any $\xi\in D(\Delta_{\psi\psi}^{-\frac s2})$,
we have
$\Delta_{\psi\psi}^{-\frac s2}\xi\in
D(\Delta_{\varphi_k\psi}^{\frac s2})$,
and 
\[
\Delta_{\varphi_k\psi}^{\frac s2}
\Delta_{\psi\psi}^{-\frac s2}\xi
=(D\varphi_k:D\psi)_{-i\frac s2}\xi,\quad  k=1,2.
\]
From this, we obtain
\begin{align}\label{pos}
\lmk T\xi,\xi\rmk=\lV(D\varphi_2:D\psi)_{-i\frac s2}\xi \rV^2
-\lV(D\varphi_1:D\psi)_{-i\frac s2}\xi \rV^2
=\lV \Delta_{\varphi_2\psi}^{\frac s2}
\Delta_{\psi\psi}^{-\frac s2}\xi\rV^2
-\lV \Delta_{\varphi_1\psi}^{\frac s2}
\Delta_{\psi\psi}^{-\frac s2}\xi\rV^2.
\end{align}
As $\Delta_{\psi\psi}^{-\frac s2}\xi$ is 
in $D(\Delta_{\varphi_2\psi}^{\frac s2})$, the last term is positive
from Lemma \ref{mono}.
This proves $T\ge 0$.\\
Next we define  $x':
=J((D\eta:D\psi)_{-i\frac{1-s}{2}})J\in{\cal M}'$.
From $\lV x'\rV\le 1$
and $0\le T\in{\cal M}$, we have
\begin{align}\label{tx}
\lmk {x'}^*Tx'\xi_\psi,\xi_\psi\rmk
=\lmk T^{\frac 12}{x'}^*x'T^{\frac 12}\xi_\psi,\xi_\psi\rmk
\le \lmk T\xi_\psi,\xi_\psi\rmk.
\end{align}
As $\xi_\psi\in D\lmk \Delta_{\psi\psi}^{-\frac{1-s}{2}}\rmk$,
from Lemma \ref{cos}, we have
\begin{align}
x'\xi_\psi
=J(D\eta:D\psi)_{-i\frac{1-s}{2}}\xi_\psi
=J\Delta_{\eta\psi}^{\frac12-\frac s2}
\Delta_{\psi\psi}^{-\frac12+\frac s2}\xi_\psi
=J\Delta_{\eta\psi}^{\frac12-\frac s2}\xi_\psi
=\Delta_{\psi\eta}^{\frac s2}
J\Delta_{\eta\psi}^{\frac12}\xi_\psi
=\Delta_{\psi\eta}^{\frac s2}\xi_\eta
\in D(\Delta_{\psi\eta}^{-\frac s2}).
\end{align}
By this and Lemma \ref{cos},
we have $\Delta_{\psi\eta}^{-\frac s2}x'\xi_\psi
\in D(\Delta_{\varphi_k\eta}^{\frac s2})$
and 
\begin{align}
(D\varphi_k:D\psi)_{-i\frac s2}x'\xi_\psi
=\Delta_{\varphi_k\eta}^{\frac s2}
\Delta_{\psi\eta}^{-\frac s2}x'\xi_\psi
=\Delta_{\varphi_k\eta}^{\frac s2}
\xi_\eta
\end{align}
for $k=1,2$.
Hence we obtain
\begin{align}\label{txl}
\lmk {x'}^*Tx'\xi_\psi,\xi_\psi\rmk
%=\lV
%\Delta^{\frac s2}_{\varphi_2\eta}\Delta_{\psi\eta}^{-\frac s2}
%J\Delta_{\eta\psi}^{\frac 12-\frac s2}\xi_\psi,
%\rV^2
%-
%\lV
%\Delta^{\frac s2}_{\varphi_1\eta}
%\Delta_{\psi\eta}^{-\frac s2}
%J\Delta_{\eta\psi}^{\frac 12-\frac s2}\xi_\psi,
%\rV^2\\
=\lV\Delta_{\varphi_2\eta}^{\frac s2}\xi_\eta\rV^2
-\lV\Delta_{\varphi_1\eta}^{\frac s2}\xi_\eta\rV^2.
\end{align}
On the other hand, substituting $\xi=\xi_\psi\in 
D(\Delta_{\psi\psi}^{-\frac s2})$ to (\ref{pos}),
we have
\begin{align}\label{tu}
\lmk T\xi_\psi,\xi_\psi\rmk
=\lV \Delta_{\varphi_2\psi}^{\frac s2}
\xi_\psi\rV^2
-\lV \Delta_{\varphi_1\psi}^{\frac s2}
\xi_\psi\rV^2.
\end{align}
From (\ref{tx}), (\ref{txl}), (\ref{tu}), we obtain the result
for the $\varphi_2\le\psi$ case.\\
To extend the result to a general case, we use Lemma \ref{eap}.
For any $\varepsilon>0$, we have
\begin{align}
\varepsilon\varphi_1\le \varepsilon\varphi_2
\le\psi+\varepsilon \varphi_2,\quad\eta\le \psi+\varepsilon\varphi_2.
\end{align}
Therefore, for any $0<s<1$, we have 
\begin{align*}
\lV\Delta_{\varepsilon\varphi_2,\eta}^{\frac s2}\xi_\eta\rV^2
-\lV\Delta_{\varepsilon\varphi_1,\eta}^{\frac s2}\xi_\eta\rV^2
\le
\lV\Delta_{\varepsilon\varphi_2,\psi+\varepsilon\varphi_2}^{\frac s2}
\xi_{\psi+\varepsilon\varphi_2}\rV^2
-\lV\Delta_{\varepsilon\varphi_1,\psi+\varepsilon\varphi_2}^{\frac s2}
\xi_{\psi+\varepsilon\varphi_2}\rV^2.
\end{align*}
Using relations  
$\Delta_{\varepsilon\varphi_2,\eta}^{\frac s2}=
\varepsilon^{\frac s2}\Delta_{\varphi_2,\eta}^{\frac s2}$ etc,
we have 
\begin{align*}
\lV\Delta_{\varphi_2,\eta}^{\frac s2}\xi_\eta\rV^2
-\lV\Delta_{\varphi_1,\eta}^{\frac s2}\xi_\eta\rV^2
\le
\lV\Delta_{\varphi_2,\psi+\varepsilon\varphi_2}^{\frac s2}
\xi_{\psi+\varepsilon\varphi_2}\rV^2
-\lV\Delta_{\varphi_1,\psi+\varepsilon\varphi_2}^{\frac s2}
\xi_{\psi+\varepsilon\varphi_2}\rV^2\\
=\lV\Delta_{\psi+\varepsilon\varphi_2,\varphi_2}^{\frac {1-s}2}
\xi_{\varphi_2}\rV^2
-\lV\Delta_{\psi+\varepsilon\varphi_2,\varphi_1}^{\frac {1-s}2}
\xi_{\varphi_1}\rV^2
\end{align*}
In the second line we used Lemma \ref{sos}.
Taking $\varepsilon \to 0$ and applying Lemma \ref{eap}
and Lemma \ref{sos},
we obtain the result.
$\square$\\\\
{\it Proof of Proposition \ref{main}}\\
%Because of Lemma \ref{sos}, it is enough to consider the case $0\le s\le 1$.
It is trivial for $s=0,1$.
We prove the claim for $0<s<1$.
We first consider faithful $\varphi,\eta$.
From $\pn\le\pn+(\en-\pn)_+$ and Lemma \ref{mono}, we have
%\begin{align}\label{san1}
%\lmk\lmk \Delta_{\pn,\pn}^s-\Delta_{\en,\pn}^{s}\rmk\xi_{\pn},\xi_\pn\rmk
%\le \lmk\lmk \Delta_{\pn+(\en-\pn)_+,\en}^s-\Delta_{\en,\pn}^{s}\rmk\xi_{\pn},\xi_\pn\rmk.
%\end{align}
\begin{align}\label{san1}
\lV\Delta_{\pn,\pn}^{\frac s2}\xi_{\pn}\rV^2
-\lV\Delta_{\en,\pn}^{{\frac s2}}
\xi_{\pn}\rV^2
\le \lV\Delta_{\pn+(\en-\pn)_+,\varphi}^{\frac s2}\xi_{\pn}
\rV^2-\lV\Delta_{\en,\pn}^{{\frac s2}}\xi_{\pn}
\rV^2.
\end{align}
By Lemma \ref{ichi} and inequalities
$\eta\le\pn+(\en-\pn)_+,\; \pn\le \pn+(\en-\pn)_+$,
the last term is bounded as
\begin{align}
\le
\lV
\Delta^{\frac s2}_{\pn+(\eta-\pn)_+,\pn+(\eta-\pn)_+}
\xi_{\pn+(\eta-\pn)_+}
\rV^2-
\lV
\Delta^{\frac s2}_{\eta,\pn+(\eta-\pn)_+}
\xi_{\pn+(\eta-\pn)_+}
\rV^2
\nonumber\\
=\lV
\xi_{\pn+(\eta-\pn)_+}
\rV^2
-
\lV
\Delta^{\frac s2}_{\eta,\pn+(\eta-\pn)_+}
\xi_{\pn+(\eta-\pn)_+}\rV^2\nonumber\\
=
\lV
\xi_{\pn+(\eta-\pn)_+}
\rV^2
-
\lV
\Delta^{\frac{1-s}{2}}_{\pn+(\eta-\pn)_+,\eta}
\xi_{\eta}\rV^2.\label{jn}
\end{align}
By $\varphi+(\eta-\pn)_+\ge \eta$ and Lemma \ref{mono},
we have
\begin{align}\label{sanyon}
(\ref{jn})\le
\lV
\xi_{\pn+(\eta-\pn)_+}
\rV^2
-
\lV
\Delta^{\frac{1-s}{2}}_{\eta,\eta}
\xi_{\eta}\rV^2
=\pn(1)+(\eta-\pn)_+(1)-\eta(1).
\end{align}
Hence
we obtain
\begin{align}
\pn(1)-
\lV\Delta_{\en,\pn}^{\frac s2}\xi_{\pn}\rV^2
\le 
\pn(1)+(\eta-\pn)_+(1)-\eta(1),
\end{align}
which is equal to
\begin{align}
\eta(1)-(\eta-\pn)_+(1)
\le 
\lV\Delta_{\en,\pn}^{\frac s2}\xi_{\pn}\rV^2.
\end{align}
\\
We now prove the inequality for general $\pn,\eta$.
By considering a von Neumann algebra $\cm_e:=e\cm e$
with $e:=s(\eta)\bigvee s(\pn)$ instead of $\cm$ if it is necessary, we may assume 
$\pn+\varepsilon \eta,\eta+\delta\pn$
are faithful on ${\cm}$ for any $\varepsilon,\delta>0$.
We then have
\begin{align}\label{appl}
\lmk \eta+\delta\pn\rmk(1)-\lmk
\eta+\delta\pn-(\pn+\varepsilon \eta)\rmk_+(1)
\le 
\lV\Delta_{\en+\delta\pn,\pn+\varepsilon \eta}^{\frac s2}\xi_{\pn+\varepsilon \eta}\rV^2.
\end{align}
Taking the limit $\varepsilon\to 0$ and then the limit $\delta\to 0$,
and using Lemma \ref{eap} and Lemma \ref{sos} we obtain
the inequality (\ref{mi}) for general $\pn,\eta$.\\
To check the condition for the equality,
by approximating $\pn$ and $\eta$
by $\pn+\varepsilon \eta,\eta+\delta\pn$
in (\ref{san1}), (\ref{jn}), and (\ref{sanyon}),
just as in (\ref{appl}), and taking 
the limit $\varepsilon\to 0$ and $\delta\to 0$,
we obtain
\begin{align}
&\lV\Delta_{\pn,\pn}^{\frac s2}\xi_{\pn}\rV^2
-\lV\Delta_{\en,\pn}^{{\frac s2}}
\xi_{\pn}\rV^2\le \lV\Delta_{\pn+(\en-\pn)_+,
\varphi}^{\frac s2}\xi_{\pn}
\rV^2-
\lV\Delta_{\en\pn}^{{\frac s2}}
\xi_{\pn}
\rV^2\nonumber\\
&\le
\lV
\Delta^{\frac s2}_{\pn+(\eta-\pn)_+,\pn+
(\eta-\pn)_+}
\xi_{\pn+(\eta-\pn)_+}
\rV^2
\quad-
\lV
\Delta^{\frac s2}_{\eta,\pn+(\eta-\pn)_+}
\xi_{\pn+(\eta-\pn)_+}
\rV^2\nonumber\\
%&=\lV
%\xi_{(\pn+\varepsilon\eta)+(\eta+\delta\pn-(\pn+\varepsilon\eta))_+}
%\rV^2\\
%&\quad-
%\lV
%\Delta^{\frac s2}_{\eta+\delta\pn,(\pn+\varepsilon\eta)+(\eta+\delta\pn-(\pn+\varepsilon\eta))_+}
%\xi_{(\pn+\varepsilon\eta)+(\eta+\delta\pn-(\pn+\varepsilon\eta))_+}\rV^2\nonumber\\
&=
\lV
\xi_{\pn+(\eta-\pn)_+}
\rV^2
-
\lV
\Delta^{\frac{1-s}{2}}_{\pn+(\eta-\pn)_+,\eta}
\xi_{\eta}\rV^2
\le
\lV
\xi_{\pn+(\eta-\pn)_+}
\rV^2
-
\lV
\Delta^{\frac{1-s}{2}}_{\eta,\eta}
\xi_{\eta}\rV^2\nonumber\\
&=\pn(1)+(\eta-\pn)_+(1)-\eta(1).
\label{long}
\end{align}
By Lemma \ref{ec}, the first inequality is an equality iff
the support of $(\en-\pn)_+$ is orthogonal to $\pn$
and the third inequality is an equality 
iff the support of $(\en-\pn)_-$ 
is orthogonal to $\en$.
Therefore, if
the equality in (\ref{long}) holds, then 
$(\en-\pn)_+$ is orthogonal to $\pn$
and $(\en-\pn)_-$ 
is orthogonal to $\en$.\\
Conversely, if $(\en-\pn)_+$ is orthogonal to $\pn$
and $(\en-\pn)_-$ 
is orthogonal to $\en$. Then
we have $\pn+(\eta-\pn)_+=\eta+(\eta-\pn)_-$,
where both sides of the equality are sum of orthogonal elements.
Therefore, we have
\begin{align}\label{nana}
&\lV
\Delta^{\frac s2}_{\pn+(\eta-\pn)_+,\pn+
(\eta-\pn)_+}
\xi_{\pn+(\eta-\pn)_+}
\rV^2
\quad-
\lV
\Delta^{\frac s2}_{\eta,\pn+(\eta-\pn)_+}
\xi_{\pn+(\eta-\pn)_+}
\rV^2=
\lV
\Delta^{\frac s2}_{(\eta-\pn)_-,\pn+(\eta-\pn)_+}
\xi_{\pn+(\eta-\pn)_+}
\rV^2\nonumber\\
&=\lV
\Delta^{\frac {1-s}2}_{\pn+(\eta-\pn)_+,(\eta-\pn)_-}
\xi_{(\eta-\pn)_-}
\rV^2
=\lV
\Delta^{\frac {1-s}2}_{\pn,(\eta-\pn)_-}
\xi_{(\eta-\pn)_-}
\rV^2
+\lV
\Delta^{\frac {1-s}2}_{(\eta-\pn)_+,(\eta-\pn)_-}
\xi_{(\eta-\pn)_-}
\rV^2\nonumber\\
&=\lV
\Delta^{\frac {1-s}2}_{\pn,(\eta-\pn)_-}
\xi_{(\eta-\pn)_-}
\rV^2=
\lV
\Delta^{\frac {s}2}_{(\eta-\pn)_-,\pn}
\xi_{\pn}
\rV^2.
\end{align}
Furthermore, we have
\begin{align*}
\lV\Delta_{\pn+(\en-\pn)_+,
\varphi}^{\frac s2}\xi_{\pn}
\rV^2-
\lV\Delta_{\en\pn}^{{\frac s2}}
\xi_{\pn}
\rV^2
=\lV\Delta_{(\en-\pn)_-,
\varphi}^{\frac s2}\xi_{\pn}
\rV^2.
\end{align*}
Hence, the second inequality is an equality in this case.
As the first and third inequalities are equalities from
the orthogonality of $(\en-\pn)_+$ with $\pn$
and $(\en-\pn)_-$ with $\en$ respectively,
the equality holds in (\ref{long}).\\
Therefore, the equality in (\ref{long}) holds iff 
$(\en-\pn)_+$ is orthogonal to $\pn$
and $(\en-\pn)_-$ 
is orthogonal to $\en$.
However, the latter condition means 
$\eta=(\en-\pn)_++\psi$ and $\pn=(\en-\pn)_-+\psi$
for some $\psi\in{\cal M}_{*+}$ whose support is orthogonal to the 
support of $\lv \en-\pn\rv$.
$\square$\\\\
{\it Proof of Corollary \ref{fcc}}\\
Replacing $\eta,\pn,s$ in (\ref{mi}) with
 $\pn,\eta,1-s$ respectively, we obtain 
\begin{align}\label{sw}
\pn(1)-(\pn-\en)_+(1)
\le 
\lV\Delta_{\pn,\en}^{\frac{1-s}{2}}\xi_{\en}\rV^2
=\lV\Delta_{\en,\pn}^{\frac{s}{2}}\xi_{\pn}\rV^2.
\end{align}
Summing (\ref{mi}) and (\ref{sw}), we obtain (\ref{mainineq}).
$\square$
\appendix
\section{Appendix}
Let $\{\cm, \ch, J,{\cal P}\}$ be a standard form associated with
a von Neumann algebra $\cm$, i.e.,
$\ch$ is a Hilbert space where $\cm$ acts on,
$J$ is the modular conjugation, and $\cal P$ is the natural positive
cone. Let ${\cal M}_{*+}$ be
the set of all positive normal linear functionals
over $\cal M$.
For each $\pn\in {\cal M}_{*+}$, $\xi_\pn$
is the unique element in the natural positive cone $\cal P$
which satisfies $\pn(x)=(x\xi_{\pn},\xi_{\pn})$
for all $x\in{\cal M}$.
For $\varphi,\psi\in {\cal M}_{*+}$,
we define an operator $S_{\pn\psi}$ as the closure of
the operator
\[
S_{\pn\psi}\lmk x\xi_\psi+(1-j(s(\psi)))\zeta\rmk:=s(\psi)x^*\xi_\pn,\quad
x\in\cm,\;\zeta\in\ch,
\]
where $s(\psi)\in{\cm}$ is the support projection of $\psi$
and $j(y):=JyJ$.
The polar decomposition of $S_{\pn\psi}$ is given by
$S_{\pn\psi}=J\Delta_{\pn\psi}^{\frac 12}$ 
where $\Delta_{\pn\psi}$
is the relative modular operator associated with
 $\varphi,\psi\in {\cal M}_{*+}$.
The subspace $\cm\xi_\psi+(1-j(s(\psi))){\cal H}$ of $\ch$
is a core of
$\Delta_{\pn\psi}^{\frac 12}$. 
The support projection of the positive operator
 $\Delta_{\pn\psi}$ is $s(\pn)j(s(\psi))$.
For a complex number $z\in{\mathbb C}$,
we define a closed operator $\Delta_{\pn\psi}^z$
by
\[
\Delta_{\pn\psi}^z:=\lmk \exp\left[ z\lmk
\log \Delta_{\pn\psi}\rmk s(\pn)j(s(\psi)) \right]\rmk
s(\pn)j(s(\psi)).
\]
For an operator $A$ on a Hilbert space $\cal H$, we denote
by $D(A)$ its domain.
\begin{lem}\label{cos}
Let $\varphi,\psi$ be faithful normal positive
linear functionals over a von Neumann algebra $\cal M$.
Suppose that there exists a constant $\lambda>0$
such that $\lambda\varphi\le \psi$.
Then the cocyle ${\mathbb R}\ni t\mapsto (D\varphi:D\psi)_t\in{\cal M}$
has an extension $(D\varphi:D\psi)_z\in{\cal M}$ analytic on 
$I_{-\frac12}:=\{z\in{\mathbb C\;:\; -\frac12<\Im z<0}\}$
and bounded continuous on $\overline{I_{-\frac12}}$
with the bound $\lV (D\varphi:D\psi)_z\rV\le \lambda^{\Im z}$
for all $z\in \overline{I_{-\frac12}}$.
Furthermore, for any faithful $\zeta\in {\cal M_{*+}}$,
$0< s< \frac12$, and
any element $\xi$ in $D(\Delta_{\psi\zeta}^{-s})$,
$\Delta_{\psi\zeta}^{-s}\xi$ is in the domain of 
$\Delta_{\varphi\zeta}^{s}$, and
\begin{align}\label{se}
\Delta_{\varphi\zeta}^{s}\Delta_{\psi\zeta}^{-s}\xi
=(D\varphi:D\psi)_{-is}\xi.
\end{align}
\end{lem}
{\it Proof}
The existence and boundedness of $(D\varphi:D\psi)_z$
is proven in \cite{ararent}.
To show the latter part of the Lemma,
let $\zeta\in {\cal M_{*+}}$ be faithful.
We define the region $I_{-s}$ in the complex plane
by $I_{-s}:=\{z\in{\mathbb C\;:\; -s<\Im z<0}\}$
for each $0< s< \frac12$.
For any $\xi\in D(\Delta_{\psi\zeta}^{-s})$ and
$\xi_1\in D(\Delta_{\varphi\zeta}^{s})$,
we consider two functions on $\overline{I_{-s}}$ by
$F(z):=\lmk\Delta_{\psi\zeta}^{-iz}\xi,
\Delta_{\varphi\zeta}^{-i\bar z}\xi_1\rmk$,
and $G(z):=\lmk (D\varphi:D\psi)_z\xi,\xi_1\rmk$.
Both of these functions are bounded continuous on 
$\overline{I_{-s}}$ and analytic on $I_{-s}$.
Furthermore, they are equal on $\mathbb R$:
\begin{align*}
F(t)=\lmk\Delta_{\varphi\zeta}^{it}
\Delta_{\psi\zeta}^{-it}\xi,
\xi_1\rmk
=\lmk(D\varphi:D\psi)_t
\xi,
\xi_1\rmk=G(t),\quad\forall t\in{\mathbb R}.
\end{align*}
This means $F(z)=G(z)$ for all $z\in \overline{I_{-s}}$.
In particular, we have $F(-is)=G(-is)$, i.e.,
\begin{align*}
\lmk\Delta_{\psi\zeta}^{-s}\xi,
\Delta_{\varphi\zeta}^{s}\xi_1\rmk=
\lmk (D\varphi:D\psi)_{-is}\xi,\xi_1\rmk.
\end{align*}
As this holds for all $\xi_1\in D(\Delta_{\varphi\zeta}^{s})$,
$\Delta_{\psi\zeta}^{-s}\xi$ is in the domain of 
$\Delta_{\varphi\zeta}^{s}$, and (\ref{se}) holds.
$\square$
\begin{lem}\label{mono}
Let $\varphi,\eta,\psi$ be normal positive 
linear functionals over a von Neumann algebra $\cal M$
such that $\varphi\le \eta$.
Then for any $0\le s\le 1$, we have
$D(\Delta_{\eta,\psi}^{\frac s2})\subset
D(\Delta_{\varphi,\psi}^{\frac s2})$
and
\begin{align}
\lV\Delta_{\varphi,\psi}^{\frac s2}\xi\rV
\le\lV\Delta_{\eta,\psi}^{\frac s2}\xi\rV,\quad \forall \xi
\in D(\Delta_{\eta,\psi}^{\frac s2}).
\end{align}
\end{lem}
{\it Proof}
This is proven in \cite{aramas}.
$\square$
\begin{lem}\label{eap}
Let $\varphi$ and $\eta$ be elements in ${\cal M}_{*+}$
and $\pn_{n}$ a sequence in ${\cal M}_{*+}$ such that 
$\lim_{n\to\infty}\lV \pn_n-\pn\rV=0$.
Then for any  and $0<s<1$,
\[
\lim_{n\to \infty}\lV
\Delta_{\varphi_n,\eta}^{\frac s2}\xi_\eta
\rV
=\lV
\Delta_{\varphi\eta}^{\frac s2}\xi_\eta
\rV.
\]
\end{lem}
{\it Proof}
By the integral representation of $t^s$,
we have
\begin{align}
\lV
\Delta_{\varphi_n,\eta}^{\frac s2}\xi_\eta
\rV^2
-\lV
\Delta_{\varphi\eta}^{\frac s2}\xi_\eta
\rV^2\nonumber\\
=\frac{\sin s\pi}{\pi}
\int_0^\infty d\lambda \lambda^{s-1}
\lmk
\lmk
\Delta_{\varphi_n,\eta}
\lmk\Delta_{\varphi_n,\eta}+\lambda\rmk^{-1}
-\Delta_{\varphi,\eta}
\lmk\Delta_{\varphi,\eta}+\lambda\rmk^{-1}
\rmk\xi_\eta,\xi_\eta
\rmk.
\end{align}
We denote the term inside of the integral by 
$f_n(\lambda)$.
It is easy to see
\begin{align*}
&\lv f_n(\lambda)\rv
\le \lambda^{s-1}\eta(1),\\
&\lv f_n(\lambda)\rv
\le \lambda^{s-2}\lmk
\lV \Delta_{\varphi_n,\eta}^{\frac 12}\xi_\eta\rV^2
+\lV \Delta_{\varphi\eta}^{\frac 12}\xi_\eta\rV^2
\rmk\le  \lambda^{s-2}\lmk
\varphi(1)+\sup_{n}\pn_n(1)
\rmk.
\end{align*}
Hence $\lv f_n(\lambda)\rv$ is bounded from above
by an integrable function independent of $n$.\\
Next we show $\lim_{n \to \infty} f_n(\lambda)=0$
for all $\lambda>0$.
To do so, we first observe that 
$\Delta_{\varphi_n,\eta}^{\frac 12}$
converges to $\Delta_{\varphi\eta}^{\frac 12}$
 in the strong resolvent sense:
For all $x\xi_\eta+(1-j(s(\eta)))\zeta
\in {\cal M}\xi_\eta+(1-j(s(\eta)))\ch$, 
using Powers-St\o rmer inequality,
we have
\begin{align*}
\lV
\Delta_{\varphi_n,\eta}^{\frac 12}
\lmk x\xi_\eta+(1-j(s(\eta)))\zeta\rmk
-\Delta_{\varphi,\eta}^{\frac 12}
\lmk
x\xi_\eta+(1-j(s(\eta)))\zeta\rmk
\rV^2
=\lV s(\eta)
x^*\xi_{\varphi_n}-s(\eta)x^*\xi_\varphi
\rV^2\\
\le
\lV x^*\rV^2\lV \xi_{\varphi_n}-\xi_\varphi\rV^2
\le
\lV x^*\rV^2\lV
\varphi_n-\varphi
\rV
\to 0,\;\;{\rm as}\;\;n\to \infty.
\end{align*}
As ${\cal M}\xi_\eta+(1-j(s(\eta)))\ch$ is a common
core for all  $\Delta_{\varphi_n,\eta}^{\frac 12}$
and $\Delta_{\varphi\eta}^{\frac 12}$, this means
$\Delta_{\varphi_n,\eta}^{\frac 12}$
converges to $\Delta_{\varphi\eta}^{\frac 12}$
 in the strong resolvent sense.
Therefore, for a bounded continuous function $g(t)=t^2(t^2+\lambda)^{-1}$,
$g(\Delta_{\varphi_n,\eta}^{\frac 12})$
converges to $g(\Delta_{\varphi\eta}^{\frac 12})$ strongly.
Hence we have $\lim_{n \to \infty} f_n(\lambda)=0$.\\
By the Lebesgue's theorem, we obtain the result. $\square$
\begin{lem}\label{ec}
For any $\pn,\eta\in{\cal M}_{*+}$ with $\pn\le\eta$
and $0<s<1$, 
\begin{align}\label{eqc}
\lV \Delta_{\eta\pn}^{\frac s2} \xi_\pn\rV
=\lV \xi_\pn\rV
\end{align}
if and only if $\eta-\pn$ is orthogonal to $\varphi$.
\end{lem}
{\it Proof}
First we prove if $\lV \Delta_{\eta\pn}^{\frac s2} \xi_\pn\rV
=\lV \xi_\pn\rV$, then $\eta-\pn$ is orthogonal to $\varphi$.
From Lemma \ref{mono}, for any 
$\zeta\in D(\Delta_{\eta\pn}^{-\frac s2})$,
$\Delta_{\eta\pn}^{-\frac s2}\zeta$ is in
$D(\Delta_{\pn\pn}^{\frac s2})$ and
\[
\lV
\Delta_{\pn\pn}^{\frac s2}
\Delta_{\eta\pn}^{-\frac s2}\zeta\rV
\le\lV\Delta_{\eta\pn}^{\frac s2}
\Delta_{\eta\pn}^{-\frac s2}\zeta
\rV\le\lV\zeta\rV.
\]
Therefore, $\Delta_{\pn\pn}^{\frac s2}
\Delta_{\eta\pn}^{-\frac s2}$ defined on 
$D(\Delta_{\eta\pn}^{-\frac s2})$ can be uniquely extended
to a bounded operator $A$ on $\cal H$, with norm
$\lV A\rV\le 1$.
We define an operator $0\le T\le 1$ by $T:=A^*A$.
Note that
\[
A\Delta_{\eta\pn}^{\frac s2}\xi_\pn
=\Delta_{\pn\pn}^{\frac s2}
\Delta_{\eta\pn}^{-\frac s2}\Delta_{\eta\pn}^{\frac s2}\xi_\pn
=\Delta_{\pn\pn}^{\frac s2}
s(\eta)
\xi_\pn
=\Delta_{\pn\pn}^{\frac s2}
\xi_\pn=\xi_\pn.
\]
From this, and the assumption, we have
\[
\lmk T\Delta_{\eta\pn}^{\frac s2}\xi_\pn,\Delta_{\eta\pn}^{\frac s2}\xi_\pn\rmk
=\lV A\Delta_{\eta\pn}^{\frac s2}\xi_\pn\rV^2
=\lV \xi_\pn\rV^2=\lV \Delta_{\eta\pn}^{\frac s2}\xi_\pn\rV^2.
\]
As the spectrum of $T$ is included in $[0,1]$,
this equality means 
\begin{align}\label{te}
T\Delta_{\eta\pn}^{\frac s2}\xi_\pn
=\Delta_{\eta\pn}^{\frac s2}\xi_\pn.
\end{align}
For any $\zeta\in D(\Delta_{\eta\pn}^{\frac s2})$,
we have
\begin{align}
\lmk\Delta_{\eta\pn}^{\frac s2}\xi_\pn,\Delta_{\eta\pn}^{\frac s2}\zeta\rmk
=\lmk T\Delta_{\eta\pn}^{\frac s2}\xi_\pn,
\Delta_{\eta\pn}^{\frac s2}\zeta\rmk
=\lmk A\Delta_{\eta\pn}^{\frac s2}\xi_\pn,
A\Delta_{\eta\pn}^{\frac s2}\zeta\rmk
=\lmk\xi_\pn,\zeta\rmk,
\end{align}
from (\ref{te}).
Therefore, $\xi_\pn\in D(\Delta_{\eta\pn}^{s})$
and $\Delta_{\eta\pn}^{s}\xi_\pn=\xi_\pn$.
Hence we obtain $\Delta_{\eta\pn}^{\frac 12}\xi_\pn=\xi_\pn$.
From this, we have
\[
s(\pn)\xi_\eta=J\Delta_{\eta\pn}^{\frac 12}\xi_\pn=\xi_\pn.
\]
We then obtain
\[
\lmk\eta-\pn\rmk(s(\pn))=0,
\]
i.e., the support of $\eta-\pn$ is orthogonal to 
the support of $\varphi$.\\
Conversely, if the support of $\eta-\pn$ is orthogonal to $\varphi$,
then we have
\begin{align}
\lV \Delta_{\eta\pn}^{\frac s2} \xi_\pn\rV^2
=\lV \Delta_{\eta-\pn,\pn}^{\frac s2} \xi_\pn\rV^2
+\lV \Delta_{\pn\pn}^{\frac s2} \xi_\pn\rV^2
=\lV \xi_\pn\rV^2.
\end{align}
$\square$
\begin{lem}\label{sos}
For all normal positive linear functionals $\psi_1,\psi_2$ over
a von Neumann algebra $\cal M$,
and $0\le s\le 1$,
\begin{align}
\lV \Delta_{\psi_1,\psi_2}^{\frac s2}\xi_{\psi_2}\rV
=\lV \Delta_{\psi_2,\psi_1}^{\frac{1-s}2}\xi_{\psi_1}\rV.
\end{align}
\end{lem}
{\it Proof} Functions $F(z):=\lmk\Delta_{\psi_1,\psi_2}^{\frac z2}\xi_{\psi_2}
\Delta_{\psi_1,\psi_2}^{\frac {\bar{z}}2}\xi_{\psi_2}\rmk$
and $G(z):=\lmk\Delta_{\psi_2,\psi_1}^{\frac {1- z}2}\xi_{\psi_1}
\Delta_{\psi_2,\psi_1}^{\frac {1-\bar{z}}2}\xi_{\psi_1}\rmk$
are bounded continuous on $0\le\Re z\le 1$ and
analytic on $0<\Re z< 1$.
It is easy to check $F(it)=G(it)$ for $t\in{\mathbb R}$.
Hence we obtain $F(z)=G(z)$ on $0\le\Re z\le 1$.$\square$
\noindent
{\bf Acknowledgement.}\\
The author is grateful for Professor N.Ozawa, for giving her 
the simple proof of the matrix inequality (\ref{fci}).
She also thanks for Prof. Jak\v{s}i\'c and Prof. Seiringer 
for kind advices on this note, and  Prof. Y. Kawahigashi,
Prof. H. Kosaki, and Prof. C.-A. Pillet for kind discussion.
The present research is 
supported by JSPS 
Grant-in-Aid for Young Scientists (B), 
Hayashi Memorial Foundation for Female Natural 
Scientists, Sumitomo Foundation, and Inoue Foundation.


\begin{thebibliography}{notitle}

\bibitem[A1]{ararent}
H. Araki: 
{\em Relative entropy of states of von Neumann algebras};
Pub. R.I.M.S., Kyoto Univ. {\bf 11}, 809-833, (1976).
\bibitem[AM]{aramas}
H. Araki and T.Masuda: 
{\em Positive cones and $L_p$-spaces for
von Neumann algebras};
Pub. R.I.M.S., Kyoto Univ. {\bf 18}, 339-411, (1982).
\bibitem[ACMMABV]{fc}
K.M.R. Audenaert, J. Calsamiglia, Ll. Masanes, R. Munoz-Tapia, A. Acin, E. Bagan, F. Verstraete:
{\em The Quantum Chernoff Bound};
Phys. Rev. Lett. 98, 160501 (2007)
\bibitem[JOPS]{jops}
V. Jak\v{s}i\'{c} Y.Ogata C.-A.Pillet and R. Seiringer\; In preparation
\bibitem[T]{Takesaki}
M. Takesaki:
{\em Theory of Operator Algebras II};
Springer Encyclopedia of Mathematical Sciences Vol 125 (2001).


\end{thebibliography}
\end{document}